\documentclass[final,3p,twocolumn,times]{elsarticle}
\usepackage{graphicx}
\usepackage{subfig}
\usepackage[hyphens]{url}
\usepackage{flushend}

\journal{~}

\begin{document}

\begin{frontmatter}
\title{Impact of the First Months of War on Routing and Latency in Ukraine}

\author[iit]{Valerio~Luconi}
\ead{valerio.luconi@iit.cnr.it}
\author[dip]{Alessio~Vecchio\corref{cor1}}
\ead{alessio.vecchio@unipi.it}

\cortext[cor1]{Corresponding author}
\address[dip]{Dip. di Ing. dell'Informazione, Universit\`a di Pisa, Largo L. Lazzarino 1, 56122 Pisa, Italy}
\address[iit]{Istituto di Informatica e Telematica, Consiglio Nazionale delle Ricerche, Via G. Moruzzi, 1, 56124 Pisa, Italy}



\begin{abstract}
Given the fundamental role of the Internet in our lives, a better understanding of its operational status during war times is crucial. In this paper, we analyze the Ukrainian Internet during the first months of war after the Russian invasion occurred in February 2022. The analysis is carried out from two points of view: routing and latency. In terms of routing, there is a substantial increase in BGP announcements and withdrawals which can be due to both physical unavailability of facilities and cyberattacks. Latency also increased significantly compared to the pre-conflict period, especially when considering paths going from Ukraine to Russia. The increase in latency appears to be due to a partial shift from peering to transit. 
As far as we know, this is the first quantitative analysis about the Internet of a large country during a major conflict. 
\end{abstract}


\end{frontmatter}

\section{Introduction and Related Work}

The operations of mass media, businesses, government agencies, and public safety organizations depend on Internet-based communications. Undermining the Internet of a country imposes a severe toll on the operational status of many critical sectors, due to the increasingly interconnected nature of services, communications, and also physical assets. At the same time, the Internet provides access to vital information to single individuals and it is the substrate upon which remote work is made possible. The importance of the Internet is even greater during catastrophic events, such as wars, when receiving news and coordinating activities has an impact on the safety of people. In this paper, we report our findings concerning the operational status of the Ukrainian Internet during the first two months and a half of the war. The analysis is carried out from two points of view: routing and latency. The first allows a better understanding of how the network adapted to the events happening both in the physical world, such as the massive movements of people, and unavailability of communication infrastructure, and in the digital domain, such as cyberattacks. The second allows quantifying the performance loss as perceived by end-users.

The Ukrainian-Russian conflict roots back in 2014, with the annexation of the Crimean peninsula to the Russian Federation. During the subsequent years, the Internet in Crimea was subject to radical changes in terms of connectivity and regulation, as documented in~\cite{9142776}. Two events were associated with significant changes in routing: the deployment of new cables connecting the Crimean peninsula to Russia through the Kerch strait, and a block imposed by the Ukrainian government on Crimea-originated traffic and directed to Russian social networks, mail services, and search engines. Traffic previously going through Ukraine, increasingly started to be routed through  Miranda Media, Rostelecom, Fiord, and UMLC, a set of Russia-based ISPs and transit providers~\cite{9142776}. 
The geopolitical significance of Internet routes is discussed in \cite{donbass21} with a specific focus on the Donbas region. A longitudinal analysis of the connectivity of the Autonomous Systems (ASes) located in Ukraine, revealed that the ones more closely related to the Donbas region progressively moved from the Ukrainian cyberspace to the Russian one. In the AS graph, the Donbas cluster appears, in the later years, to be placed at the periphery of the Ukrainian Internet, but still not fully integrated into the Russian one \cite{donbass21}. Some anecdotal evidence also suggested that the physical paths covered at the IP level could be radically different, depending if the source was located in the part of the territory controlled by the Ukrainian government, or in one of the separatist republics \cite{donbass21}. The destination for the two paths was always the same, Moscow, but in the first case the path was circuitous and transited through international carriers to avoid the Ukraine-Russia border, whereas the second one was more direct. Such a study, however, is lacking in terms of statistical significance as the number of analyzed paths was too low. 

The impact of the 2022 conflict on Internet traffic has been observed by Cloudflare monitoring infrastructure in the 21 Feb - 4 Mar period~\cite{graham-cumming_2022}. Several phenomena are visible in the traffic patterns: there is an increase in the level of traffic in western cities of Ukraine due to the movements of people towards the border, and a decrease in the level of traffic in cities closer or directly involved in battles. At the same time, they observed an increased number of cyberattacks, as level 3/4 and level 7 Distributed Denial of Service (DDoS), highlighting how the conflict in the real world is accompanied by hostile activities in cyberspace. The intertwined relationship between the real world and cyberspace during the conflict is testified also by the following event: Cloudflare moved the customer encryption key material out of their data centers in Ukraine, Russia, and Belarus still preserving operations via more secure data centers; in addition, the machines were configured to self-brick in case of power or connection losses~\cite{prince_2022}. 

DDoS and possible BGP hijacking events occurring in the region were also reported by the Mutually Agreed Norms on Routing Security (MANRS) initiative~\cite{siddiqui_2022}. The resiliency of the Ukrainian Internet during the first 2-3 weeks was discussed in~\cite{aben}, where the lack of market concentration, as well as the relatively large number of Internet eXchange Points (IXPs) providing connectivity to the country were found as contributing factors to the rather surprisingly tolerance of the network despite the major devastation occurred. 

The impact of some catastrophic events on the Internet was studied in the past. In \cite{10.1145/2079360.2079362}, the effects of a major earthquake in Japan were analyzed from the point of view of routing and traffic as seen from an ISP. The network outages caused by Hurricane Sandy were evaluated in \cite{Heidemann12d}. The COVID-19 pandemic also had a significant impact on the Internet in terms of traffic and latency \cite{FAVALE2020107290,CANDELA2020107495}. The role of the Internet in situations of sociopolitical turmoil was also a matter of attention, in particular concerning the revolts in Egypt that occurred in 2011 \cite{zhuo2011egypt, doi:10.1177/1748048512459147}, but in this case the focus was on the Internet as a communication technology. The effects of a potential disaster -- solar superstorms -- were also evaluated in terms of possible network outages \cite{10.1145/3452296.3472916}. The impact of potential large-scale disasters was faced also according to simulation-based approaches \cite{modelling}. 

We observed the impact of war on the Internet in Ukraine in terms of routing and latency. As far as we know, this is the first time the Internet is observed during a major conflict. The analysis spans approximately 12 weeks, the longest possible period at the time of writing considering the necessary collection, processing, and synthesis of information. 

\section{Dataset}
\label{sec:dataset}
We collected an extensive dataset about routing, latency, Internet paths, and Internet geography involving Russian and Ukrainian ASes, in the period from 14 Feb 2022 to 7 May 2022. The observation period begins ten days before the start of the Russian invasion of Ukraine (24 Feb 2022). These ten days are used as a baseline against which the observations during the war are compared. Data have been collected from multiple sources managed by RIPE NCC~\cite{ripencc}, which is the Regional Internet Registry for Europe, the Middle East and parts of Central Asia. Besides this activity, RIPE NCC also handles Internet measurement platforms, and publishes all the data free for everyone. We collected data from the following sources:

\begin{itemize}
    \item RIPE RIS~\cite{riperis}. RIPE RIS is a project that collects raw routing data from over 500 cooperating ASes via 25 route collectors. The routing data is in the form of BGP updates and BGP Routing Information Bases (RIBs). BGP is the routing protocol used to establish inter-domain routing across the whole Internet~\cite{bgprfc}. BGP updates are control messages exchanged by BGP peers to establish or remove Internet routes. In our study, we are interested in two particular messages: \textit{announcements} and \textit{withdrawals}. Announcements are used by ASes to communicate their prefixes reachability. They are propagated to establish or update a path to a prefix (route) from elsewhere in the Internet. Withdrawals are instead used to communicate that a route is no longer available. We collected all BGP updates from 14 Feb 2022 to 7 May 2022. To collect and parse BGP updates we used BGPStream, a library distributed by CAIDA, which provides an API for collecting and parsing BGP data~\cite{bgpstream}.
    \item RIPEstat~\cite{ripestat}. RIPEstat is a platform that provides an easy-to-use API for downloading aggregated data extracted from: (i) raw BGP data collected by RIPE RIS, (ii) routing and administrative data from Internet Routing Registries (IRRs), and (iii) prefix geolocation data extracted from MaxMind GeoLite databases. We collected data about all the routed ASes and their routed prefixes in Ukraine and Russia from 14 Feb 2022 to 7 May 2022. Routed ASes are ASes that are seen in BGP data collected by RIPE RIS, while their prefixes are the prefixes that are announced by them via BGP. In addition, we collected geolocation data about all the Russian and Ukrainian ASes.
    \item RIPE Atlas~\cite{staff2015ripe}. RIPE Atlas is an Internet measurement platform that performs active measurements such as Ping, Traceroute, and HTTP probing. Measurements are carried out by nodes spread all over the world. RIPE Atlas nodes belong to two categories: probes and anchors. Probes are hosted by volunteers, typically in their home networks, or in the network of small-medium companies. Anchors are more powerful nodes and are typically hosted in the networks of larger and well-connected companies or organizations such as IXPs, data centers, and operational centers of ISPs. Probes automatically carry out some network measuring tasks called Anchoring Measurements (AMs), where anchors play the role of targets. The global number of nodes belonging to the RIPE Atlas platform is approximately 12\,000 (in particular, $\sim$11\,200 probes and $\sim$760 anchors), distributed worldwide. For our study, we extracted the results of Ping, HTTP, and Traceroute AMs performed by the Atlas nodes in Ukraine from 14 Feb 2022 to 7 May 2022. The data is provided by devices scattered over a significant part of Ukraine and hosted in different ASes. As a consequence, the data used for the analysis originates from a set of vantage points that is heterogeneous and reasonably stable, and allows observing the phenomena from a country-scale perspective not tied to the specific point of view of a single stakeholder. 
    
\end{itemize}

\section{Impact on Routing}
\label{sec:routing}

\begin{figure}[!t]
    \centering
    \includegraphics[width=\columnwidth]{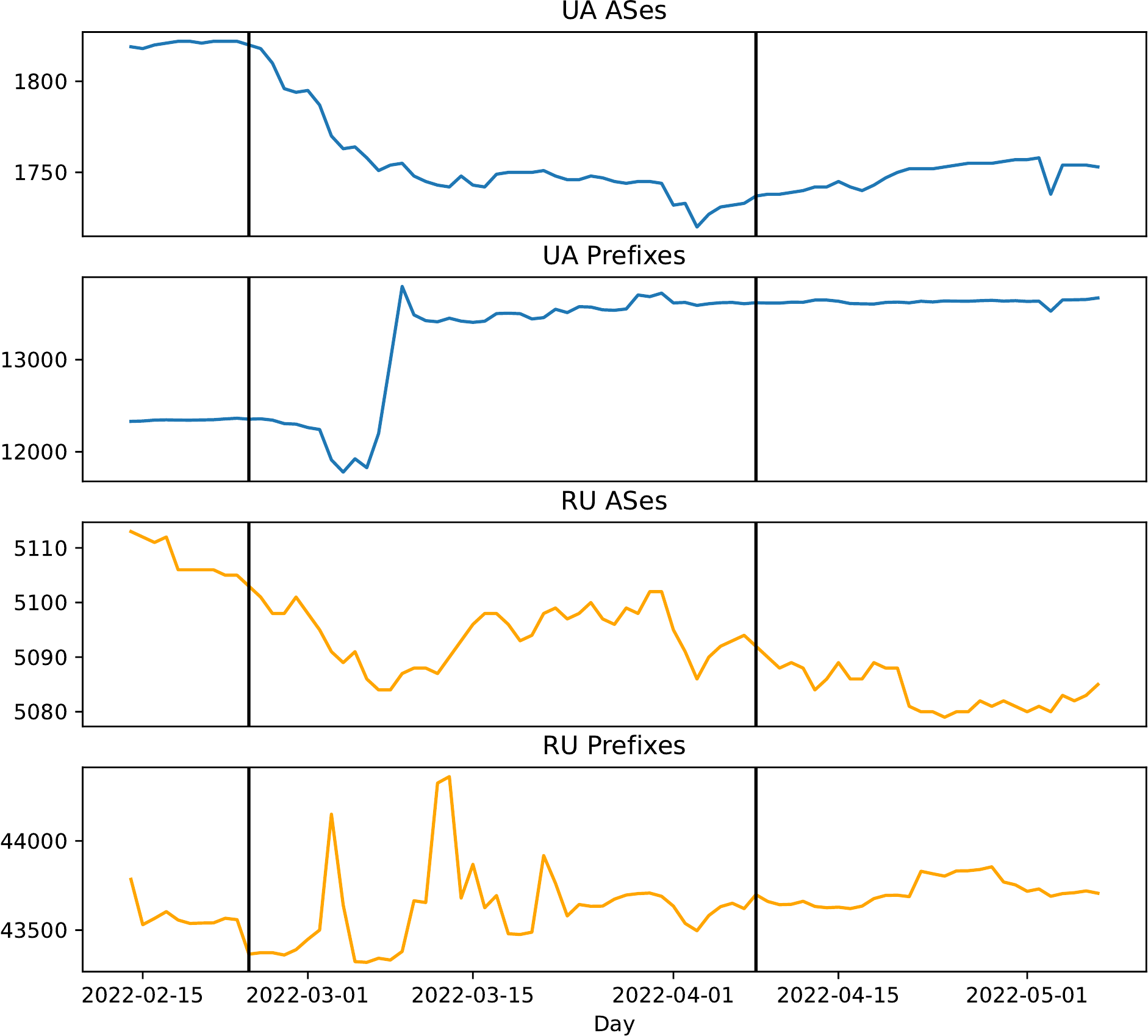}
    \caption {Number of routed ASes and routed prefixes per day for both Ukraine and Russia.}
    \label{fig:ases-prefixes}
\end{figure}

\begin{figure*}[!t]
 \centering
    \subfloat[Density of the positions of the ASes that showed some loss of connectivity.]{
        \includegraphics[width=0.9\columnwidth]{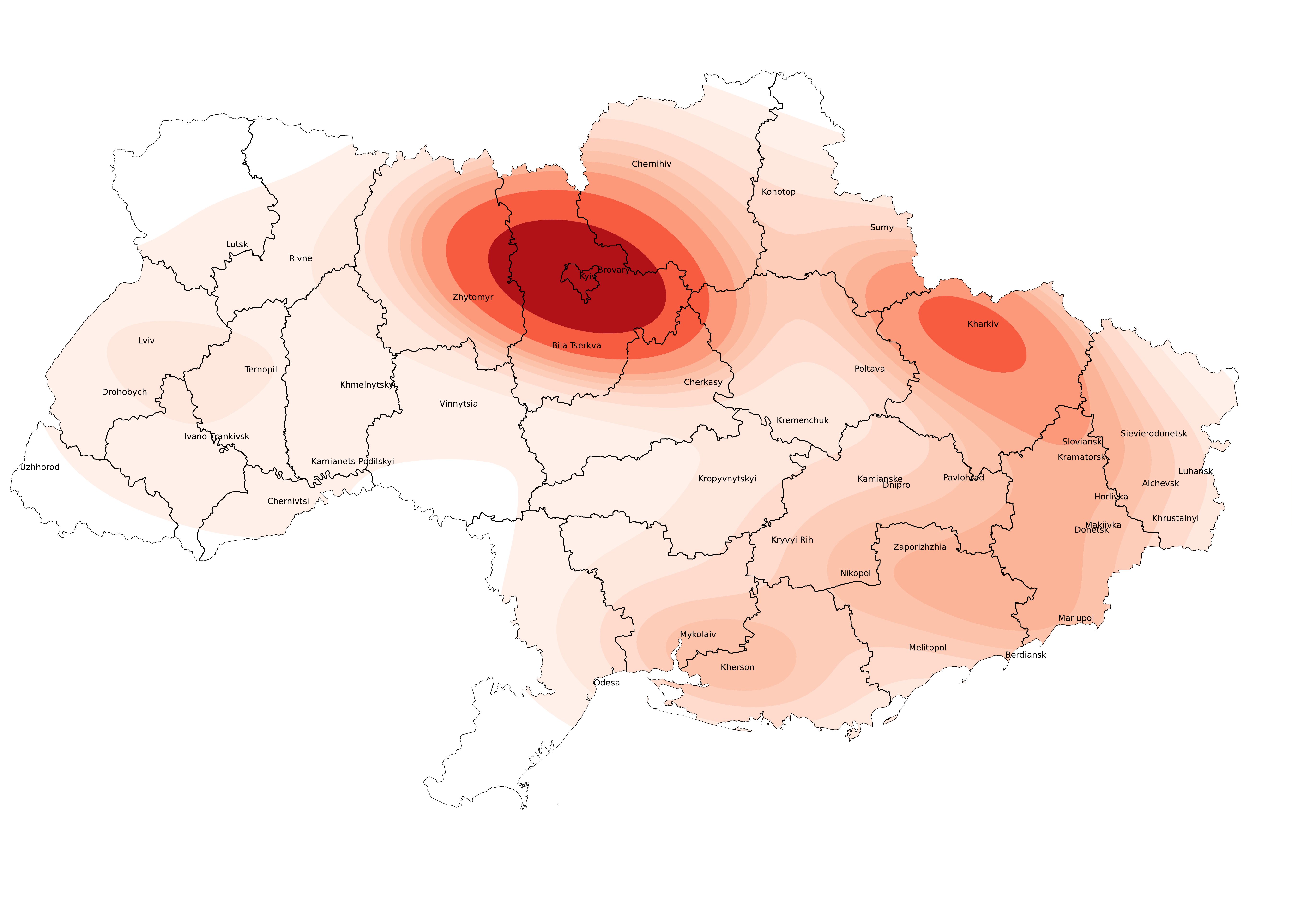}
        \label{subfig:ua-ases-map}
    }\qquad
    \subfloat[Density of the positions of Russian attacks and battles on the Ukrainian soil.]{
        \includegraphics[width=0.9\columnwidth]{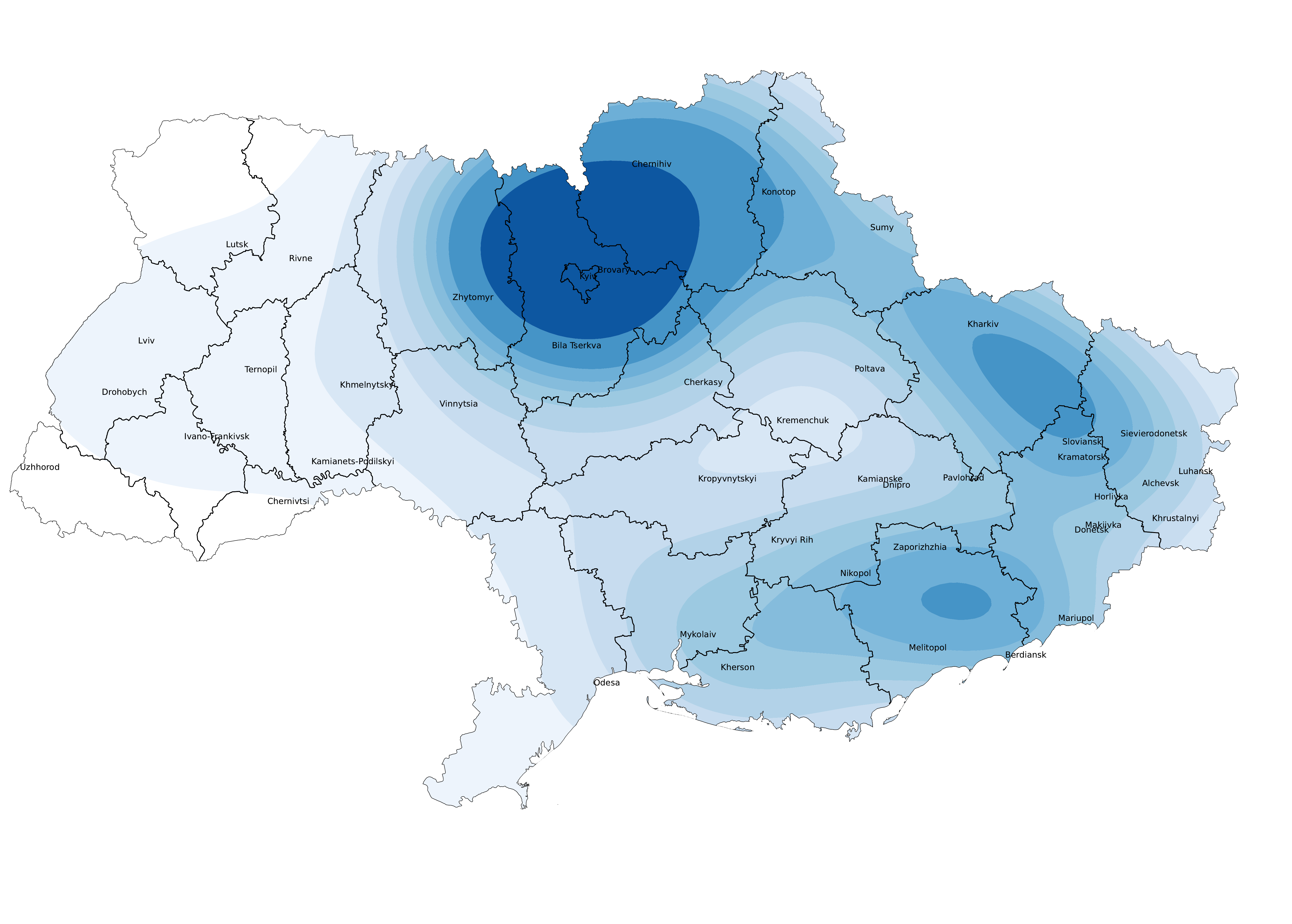}
        \label{subfig:ua-war-map}
    }
    \caption{Ukrainian ASes and war events maps.}
    \label{fig:maps}
\end{figure*}

In this section, we show the impact of the conflict on Internet routing, from the perspective of both countries.

\subsection{Routed ASes and Prefixes}
\label{subsec:ases}

Figure~\ref{fig:ases-prefixes} shows the evolution over time of the number of routed ASes and prefixes in both the Ukrainian and Russian Internet. Let us first consider the case of Ukraine. As can be observed, the number of routed ASes starts decreasing right after the start of the war, from a maximum of 1\,822 right before the start of the conflict to a minimum of 1\,720 occurring approximately around 3 Apr 2022. Around this date, the Russian forces were starting to abandon northern Ukraine to concentrate their efforts on the southeast side of the country. In fact, from the start of the second phase of the war around 8 Apr 2022, the number of routed ASes slowly increases. In the considered period some of the ASes that lost connectivity showed intermittent behavior, by going offline and back online at different times. The total number of ASes that showed some loss of connectivity is 307. Some of the reasons underlying the loss of routed ASes could include the bombings and battles that happened on Ukrainian soil, which possibly destroyed part of those ASes' infrastructure or made them shut down their services due to power unavailability or evacuation. To confirm this, we gathered from RIPEstat the geolocation of the 307 Ukrainian ASes that showed some loss of connectivity in the war period, and from Wikipedia the position of Russian attacks and battles~\cite{wikibattles}. In Figure~\ref{fig:maps}, we show the densities of the positions of ASes and the position of attacks. We can observe an evident graphic correlation between the two sets of positions. 

The number of routed prefixes starts at 12\,330, then it shows a slight decrease after the first week of the war, until a minimum of 11\,779 routed prefixes, and then a sudden increase, with a number of routed prefixes even higher than in the baseline period, reaching a maximum of 13\,796. This could happen for multiple reasons. War, as known, spans also in the cyberdomain, and, in particular, Internet prefixes can be subject to cyberattacks such as BGP hijacking. In short, a BGP hijacking attack occurs when a prefix is announced by an AS other than the owner AS, with the purpose of redirecting traffic for multiple aims, including denial of service, and traffic interception (more details about BGP hijacking attacks are provided in Section~\ref{subsec:suspect}). A way to hinder certain BGP hijacking attacks is to announce more specific sub-prefixes of the attacked prefix, as the Internet routing is based on the longest prefix match. By analyzing the data, we discovered multiple cases in which new prefixes were announced by Ukrainian ASes. First, some Ukrainian ASes simply started announcing new prefixes they never announced before. The reasons behind such behavior are inscrutable using the collected data. Possible explanations that we imagine include that these ASes moved their services to new prefixes to avoid attacks (either BGP or other kinds of cyberattacks), or they moved to new facilities due to damages. Second, some Ukrainian ASes started announcing sub-prefixes of their own prefixes. This practice has been implemented in different ways. Some ASes, like Ukrtelekom, divided all their prefixes into /24 subnets (i.e., networks with 256 IP addresses). This could allow mitigating the effect of possible BGP hijacks, as attackers should target smaller prefixes (the reader should note that a /24 network is already quite small), and a higher number of prefixes to be effective. In other cases, Ukrainian ASes started announcing /32 prefixes, i.e., single IP addresses. This practice is particularly effective to limit BGP hijacks based on more specific prefixes, as nothing is more specific than a single IP address. However, in certain cases, these prefixes still went offline after a certain amount of time. Particularly interesting is the following case. In the late hours of 8 Mar 2022 and the early hours of 9 Mar 2022, a Ukrainian AS announced $\sim$ 200 prefixes owned by Russian ASes. The duration of this phenomenon has been of few hours, and we can not be sure if this has been a deliberate tentative of BGP hijacking or if it is the result of routers' misconfiguration. We will deepen the analysis on BGP hijacking attacks between Russia and Ukraine in Section~\ref{subsec:suspect}.

For Russia, the picture is quite different. The number of routed ASes decreases over time, but in a very small percentage. The number of routed ASes per day fluctuates between approximately 5\,110 and 5\,080. The total number of ASes that showed loss of connectivity is 335, very similar to the Ukrainian number, but it represents a much smaller percentage, approximately 7\% of the total compared to 17\%. The number of prefixes also fluctuates, between approximately 43\,300 and 44\,400. Thus, the fluctuations are much less evident than in the Ukrainian case. However, a certain degree of instability is observed in the first phase of the war, with some peaks around 3 Mar 2022 and 12-13 Mar 2022. Especially in those days some Russian ASes started announcing sub-prefixes of their prefixes. For example, VDS Telecom split its prefixes into /30 and /32 prefixes in 3 Mar and 12 Mar, for an amount of over 1\,400 prefixes announced. These announcements lasted just a few hours on both days. As for the Ukrainian case, this could be an attempt to recover from BGP hijacking attacks. As mentioned above, we will deepen the analysis on this aspect in Section~\ref{subsec:suspect}.

\subsection{BGP Activity}
\label{subsec:bgp}

\begin{figure}[!t]
    \centering
    \includegraphics[width=\columnwidth]{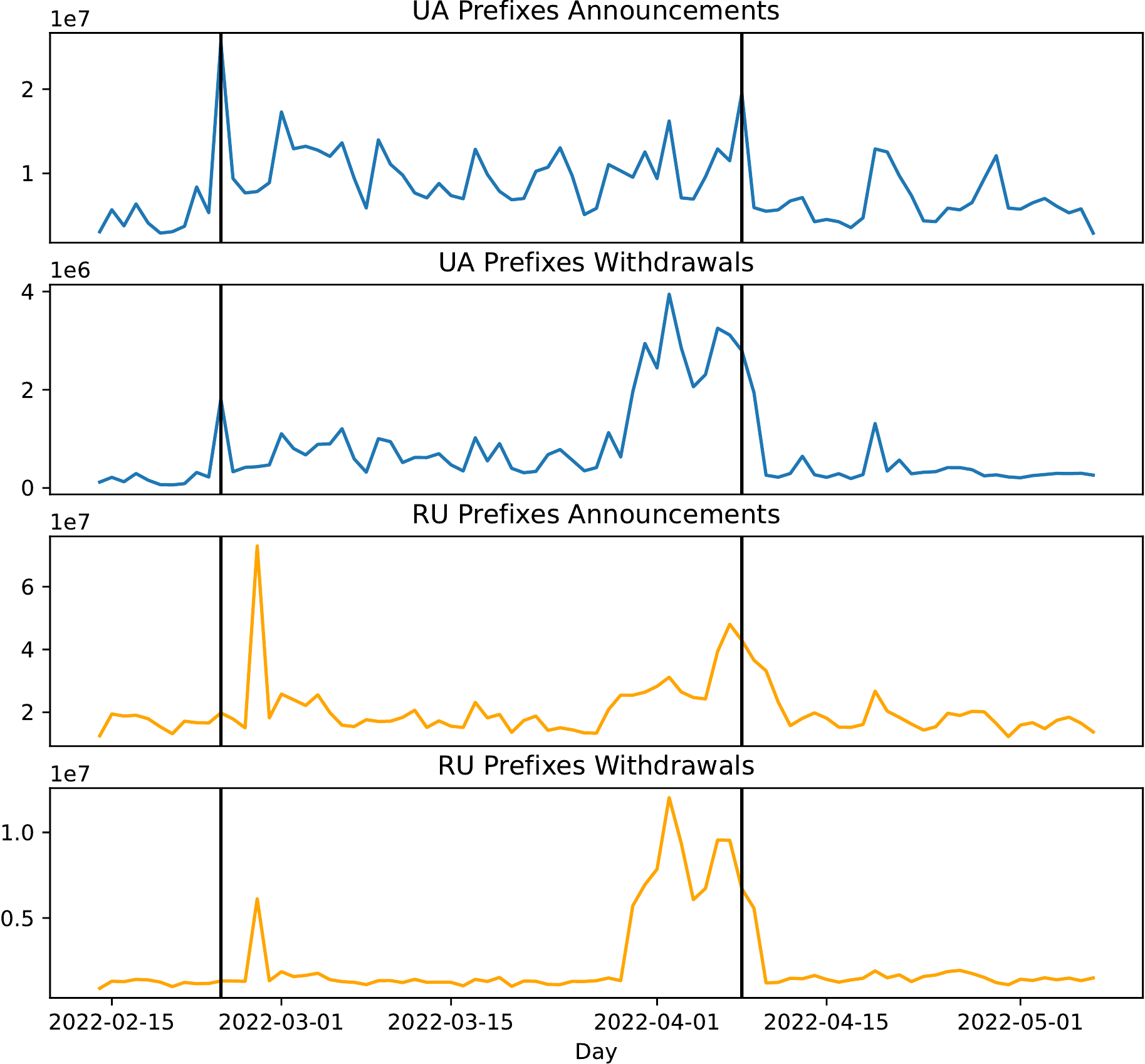}
    \caption {Number of BGP announcements and withdrawals per day for both Ukrainian and Russian prefixes.}
    \label{fig:updates}
\end{figure}

\begin{figure}[!t]
 \centering
    \subfloat[UA announcements.]{
        \includegraphics[width=\columnwidth]{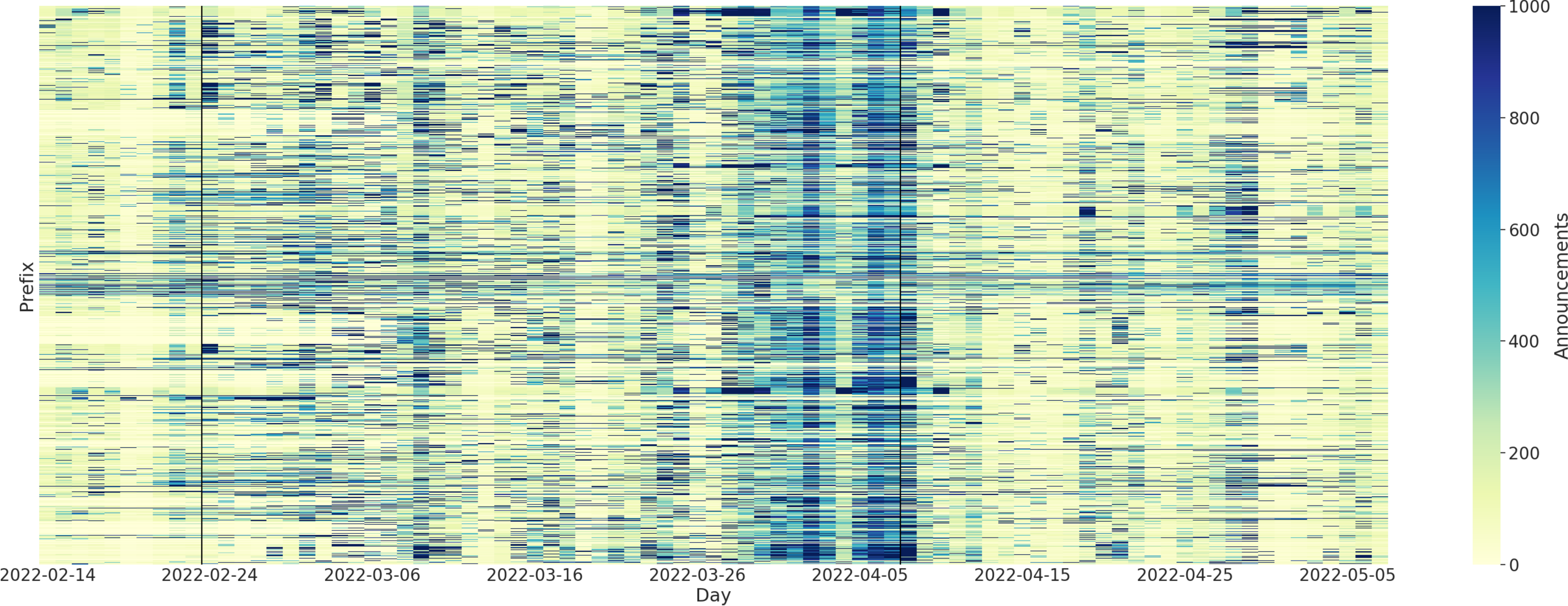}
        \label{subfig:ua-announce}
    }\\
    \subfloat[UA withdrawals.]{
        \includegraphics[width=\columnwidth]{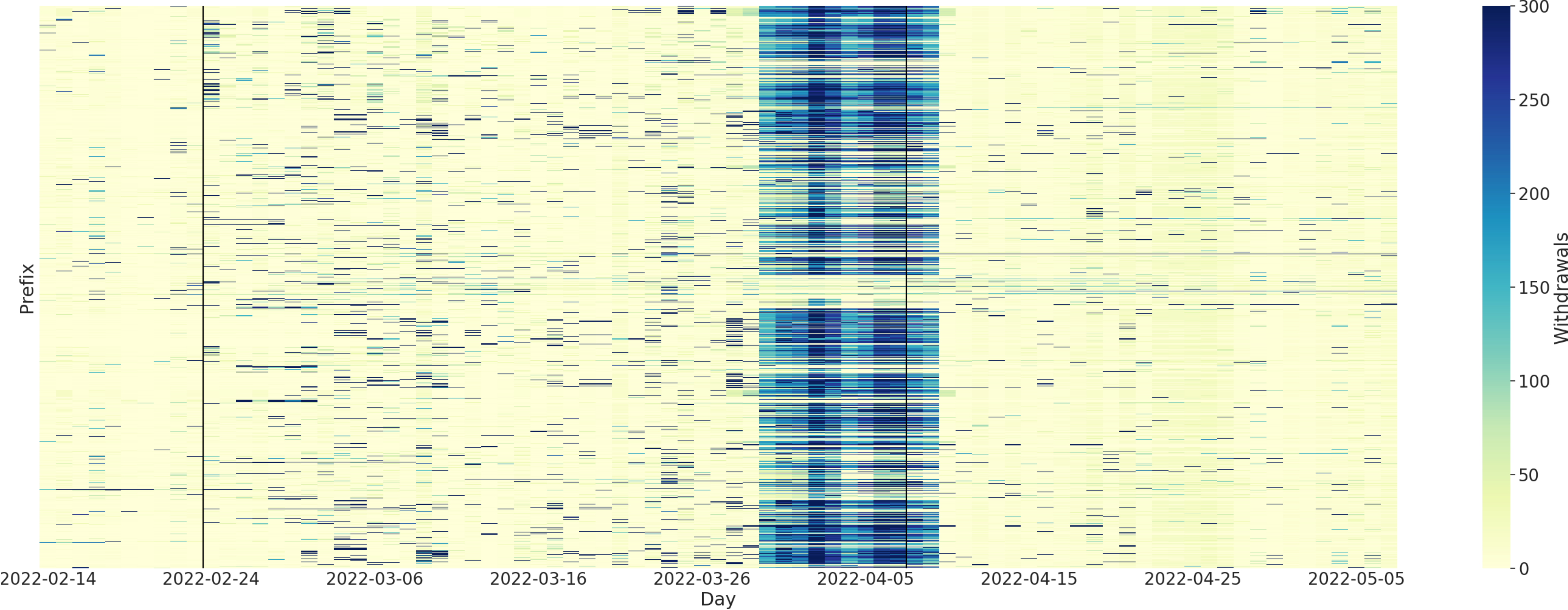}
        \label{subfig:ua-withdraw}
    }\\
    \subfloat[RU announcements.]{
        \includegraphics[width=\columnwidth]{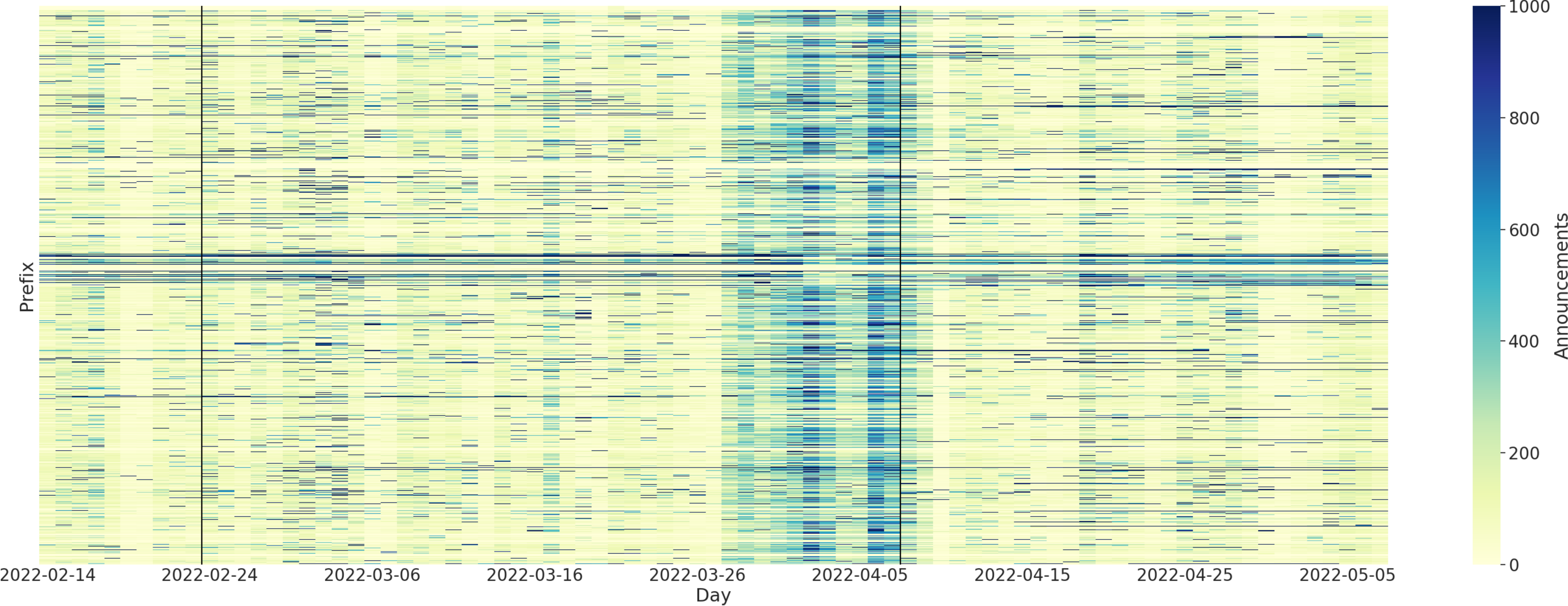}
        \label{subfig:ru-announce}
    }\\
    \subfloat[RU withdrawals.]{
        \includegraphics[width=\columnwidth]{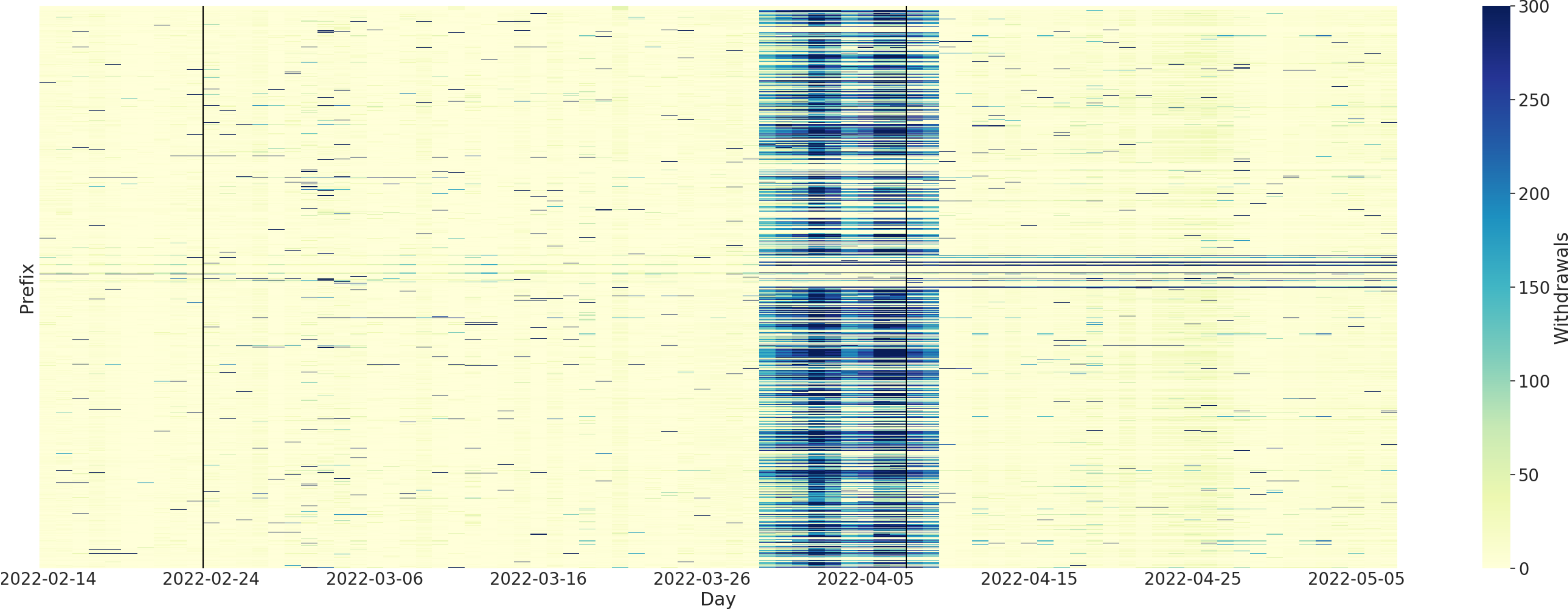}
        \label{subfig:ru-withdraw}
    }\\
    \caption{Heatmaps showing the number of BGP updates per day per prefix for both Ukrainian and Russian prefixes. Each line on the y axis represents an announced prefix, the x axis shows the time, and the color represents the number of updates, as shown in the color map on the right.}
    \label{fig:heatmaps}
\end{figure}

Figure~\ref{fig:updates} shows the number of BGP announcements and withdrawals per day for both Ukrainian and Russian prefixes. Let us first consider the Ukrainian case. As can be observed in the first plot of Figure~\ref{fig:updates}, the number of BGP announcements per day grows significantly as the Russian invasion starts. This can indicate a cyber warfare scenario as described above, or simply that the Ukrainian Internet needed to reconfigure due to damage or destruction of the infrastructure. Figure~\ref{fig:heatmaps} shows the number of BGP updates per day and per prefix for both Ukrainian and Russian prefixes. Each line on the y axes of the four subfigures represents an announced prefix, and the color represents the number of updates on a specific day. As can be observed from Figure~\ref{subfig:ua-announce}, the increment of BGP announcements is spread over the vast majority of prefixes and begins two-three days before the start of the war. However, the activity on 24 Feb 2022, which corresponds to a peak in the first plot of Figure~\ref{fig:updates}, seems to involve a limited number of prefixes, while in the period between 28 Mar 2022 and 9 Apr 2022 there is an intense activity of BGP announcements widely diffused in all prefixes, even if the total number of announcements per day is smaller than in 24 Feb 2022. Particularly interesting is the total number of BPG withdrawals per day, which in the second plot of Figure~\ref{fig:updates} shows a substantial increment in the war period, especially in the days from 28 Mar 2022 to 9 Apr 2022. On those days, an evident bump can be observed: the number of withdrawals is from 4 to 8 times higher than in every other day of the conflict. Figure~\ref{subfig:ua-withdraw} shows that the increment is spread over almost all the prefixes, with very few exceptions. On the other war days, a slightly incremented activity is visible, with peaks for a few isolated prefixes, especially during the first phase of the Russian invasion. By observing Figure~\ref{fig:heatmaps} in its entirety, we can conclude that in the period going from 28 Mar 2022 to 9 Apr 2022 there has been some event that triggered an extremely anomalous number of BGP updates in (almost) all Ukrainian ASes.

The Russian case shows some similarities but also some differences. The announcement and withdrawal activities remain approximately unchanged during the whole observation period, with two notable exceptions. Both announcements and withdrawals show a peak corresponding to 27 Feb 2022, and a bump in the days from 28 Mar 2022 to 9 Apr 2022. In those days the announcements activity almost doubles, and the withdrawal activity grows six or seven times (third and fourth plots of Figure~\ref{fig:updates}). The Russian and Ukrainian withdrawal bumps are almost overlapped, except for the absolute number of updates, which for Russia is much greater. Figure~\ref{subfig:ru-announce} shows that the announcement activity in correspondence of the first peak (27 Feb 2022) does not seem to be spread over all the prefixes. The same happens for withdrawals, as shown in Figure~\ref{subfig:ru-withdraw}. Instead, the announcement and withdrawal activity observed between 28 Mar 2022 and 9 Apr 2022 is spread over almost all prefixes, as in the Ukrainian case.

\subsection{Suspect Cases}
\label{subsec:suspect}
\begin{figure*}[!t]
    \centering
    \includegraphics[width=\textwidth]{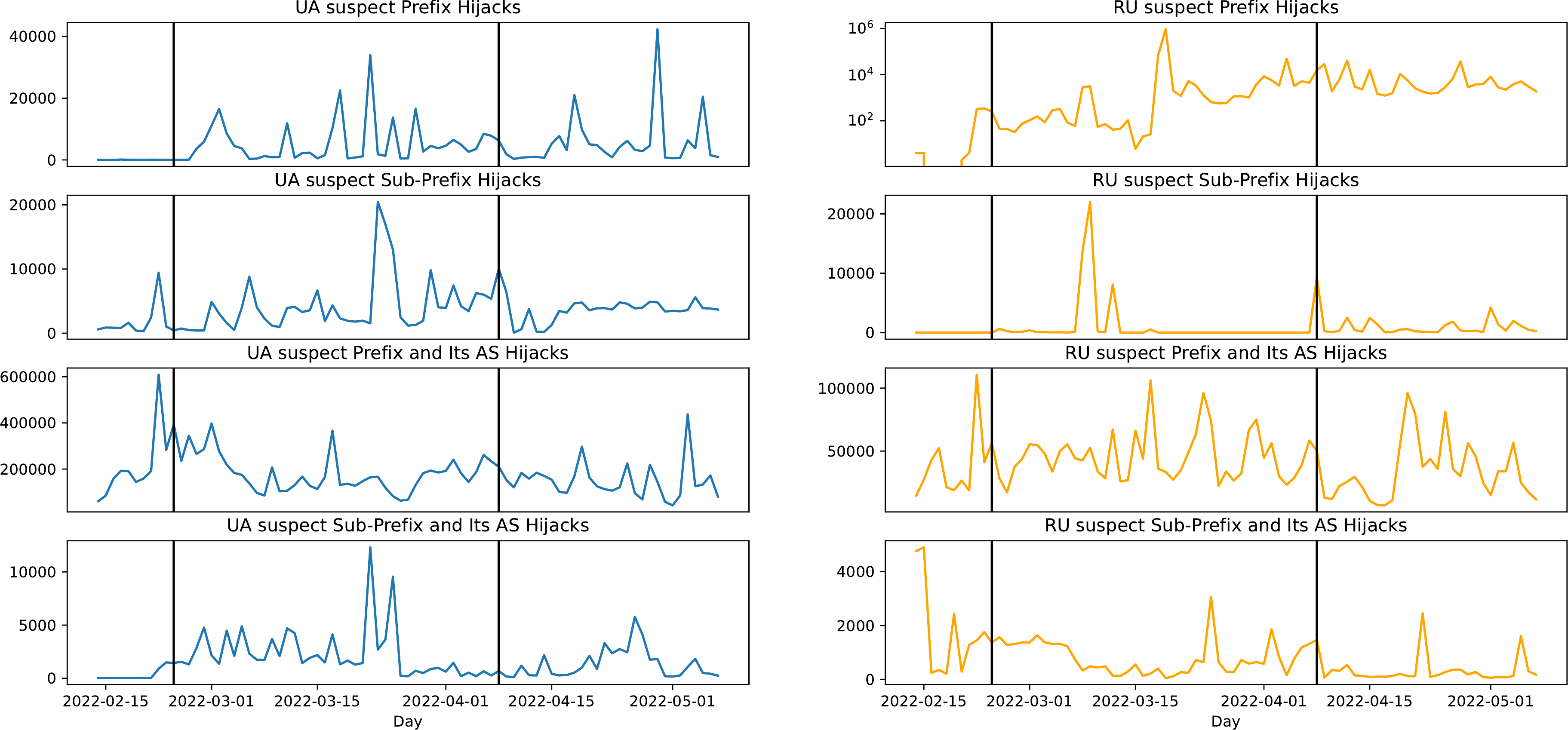}
    \caption {Number of BGP announcements showing suspect BGP hijack cases.}
    \label{fig:prefix-hijack}
\end{figure*}
In this section, we analyze BGP updates that could indicate the occurrence of BGP hijacking attacks. We rely on the classification of BGP hijacking attacks provided in~\cite{Almusawi2017:bgp}, which we report briefly. BGP hijack attacks are classified as: (i) Prefix Hijack, (ii) Prefix and Its AS Hijack, (iii) Sub-Prefix Hijack, (iv) Sub-Prefix and Its AS Hijack, (v) Hijack a Legitimate Path. In the first case, the attacker announces an existing prefix of the victim, as if it was the owner. In the second case, the attacker generates an announcement for an existing prefix of the victim as if the victim was directly connected to the attacker. In this case, the victim still appears as the owner of the prefix. The third and the fourth cases are identical to the first and second cases, except that instead of a prefix, the attacker announces a sub-prefix of an existing prefix of the victim. In the fifth and last case, the attacker propagates an existing announcement but putting itself as if it was a neighbor of the victim. In practice, this last case is almost indistinguishable from the second case, as when an announcement is collected we are unable to tell if it has been manipulated or generated from scratch.

We found some cases of suspect BGP hijacking from Russian attacker ASes to Ukrainian victim ASes, and vice-versa. To identify such possible BGP hijacking cases, as a baseline we built a table with the prefixes belonging to Ukrainian and Russian ASes collected on 14 Feb 2022 from RIPEstat, coupled with their owner AS. We checked the prefixes with data from Internet Routing Registries RIPE and RADB to be sure that they were assigned to the correct AS. Then, we collected from RIPE RIS all the BGP updates from 14 Feb 2022 to 7 May 2022, as explained in Section~\ref{sec:dataset}. We only considered BGP announcement updates. We identified suspect cases with the following procedure. From a BGP announcement we extract the announced prefix, the AS\_PATH, and, from the AS\_PATH, the origin AS. The AS\_PATH indicates the path that the announcement has traveled from the AS that originated the BGP update to the route collector, in terms of traversed ASes\footnote{The AS\_PATH indicates the route from the route collector AS to the AS that originated the BGP update.}. The origin AS is the AS that is originating the BGP update, i.e. that is announcing the prefix, and it can be found as the first AS in the AS\_PATH. We consider BGP announcements whose origin AS is one of the Russian or Ukrainian ASes previously identified. We then check the prefix against our prefix table that we previously built.

Figure~\ref{fig:prefix-hijack} shows the number of suspect cases of BGP hijack attacks, for both Russia and Ukraine, in blue those made by Russian attackers against Ukrainian victims (left) and in orange those made by Ukrainian attackers against Russian victims (right). In the first row are the suspect Prefix Hijacks, in the second the suspect Sub-Prefix Hijacks, in the third the suspect Prefix and Its AS Hijacks, and in the fourth the suspect Sub-Prefix and Its AS Hijack (four of the five previously mentioned categories).

We first consider suspect cases of Prefix Hijacks. In this case, the prefix of the announcement is contained in the table that we previously built, and owner AS and origin AS are different and from different countries (i.e., Russian origin and Ukrainian owner, and vice-versa). As can be observed from the first row of Figure~\ref{fig:prefix-hijack}, the number of suspect Prefix Hijacks is extremely low before the start of the Russian invasion (almost zero cases), and increases rapidly as the war starts, with peaks between 10\,000 and 40\,000 announcements for Ukraine, and almost 1\,000\,000 for Russia. In the case of Ukrainian attackers, the number of announcements showing suspect Prefix Hijack cases grows on average by two orders of magnitude during the first phase of the war. In the case of Russian attackers instead, the activity is intermittent. It must be noted that in both cases the suspect activity does not seem to decrease when the second phase of the war starts.

The second row of Figure~\ref{fig:prefix-hijack} shows cases of suspect Sub-Prefix Hijack from Russian attackers to Ukrainian victims and vice-versa. In this case, the prefix of the announcement is a sub-prefix of a prefix in our table, and the origin AS and the owner AS are different and from different countries. As can be observed, in the case of Ukrainian victims and Russian attackers, the activity starts a few days before the invasion, and continues all over the duration of the observation period, with peaks reaching 20\,000 announcements per day. In the case of Ukrainian attacks, the occurrence of these suspect cases is limited to a few days. In the second phase of the war, the activity seems more frequent, albeit with a limited number of announcements per day.

The suspect Prefix and Its AS Hijacks, shown in the third row of Figure~\ref{fig:prefix-hijack}, are identified in the following way. In an announcement, we check that the prefix has the correct origin, i.e. corresponding to the owner AS from our table. Then, we observe the AS\_PATH, and we identify the nationality of all the ASes in the path, as Russian, Ukrainian, or other country. If the second AS in the path (the one after the origin) is Russian for Ukrainian origin or Ukrainian for Russian origin, we flag the announcement as suspect. We then consider only announcements whose peer AS (i.e., the AS communicating with the route collector) is of a different country from the suspect attacker. Thus, if the origin AS is Ukrainian, and the suspect attacker is Russian, the peer has to be not Russian, and vice-versa. This is because if the announcement is destined to the same country of the suspect attacker, it could be just a normal path traversing that country. As can be observed, in both Ukraine and Russia, the number of suspect Sub-Prefix and Its AS Hijacks has a peak two days before the start of the war. The peak reaches approximately 600\,000 announcements for Ukrainian AS victims, and 100\,000 announcements for Russian AS victims. In both cases, the trend of these suspect attacks is intermittent during the war period. However, in the first days of the war, the activity seems more prominent for Russian suspect attacks.

We then consider the suspect Sub-Prefix and Its AS Hijacks, shown in the fourth row of Figure~\ref{fig:prefix-hijack}. We identify these announcements as in the previous case, but in this case we consider sub-prefixes of the prefixes in our table. For Ukrainian attackers and Russian victims, the activity is not so clear: the plot starts with a peak of over 4\,000 announcements, then shows some other peaks, but in general the activity seems quite marginal. Instead, in the case of Russian attackers and Ukrainian victims, the plot shows almost no activity before the start of the invasion, and a more pronounced activity during the war period, especially in the first phase of the war. 

\begin{figure}[!t]
    \centering
    \includegraphics[width=\columnwidth]{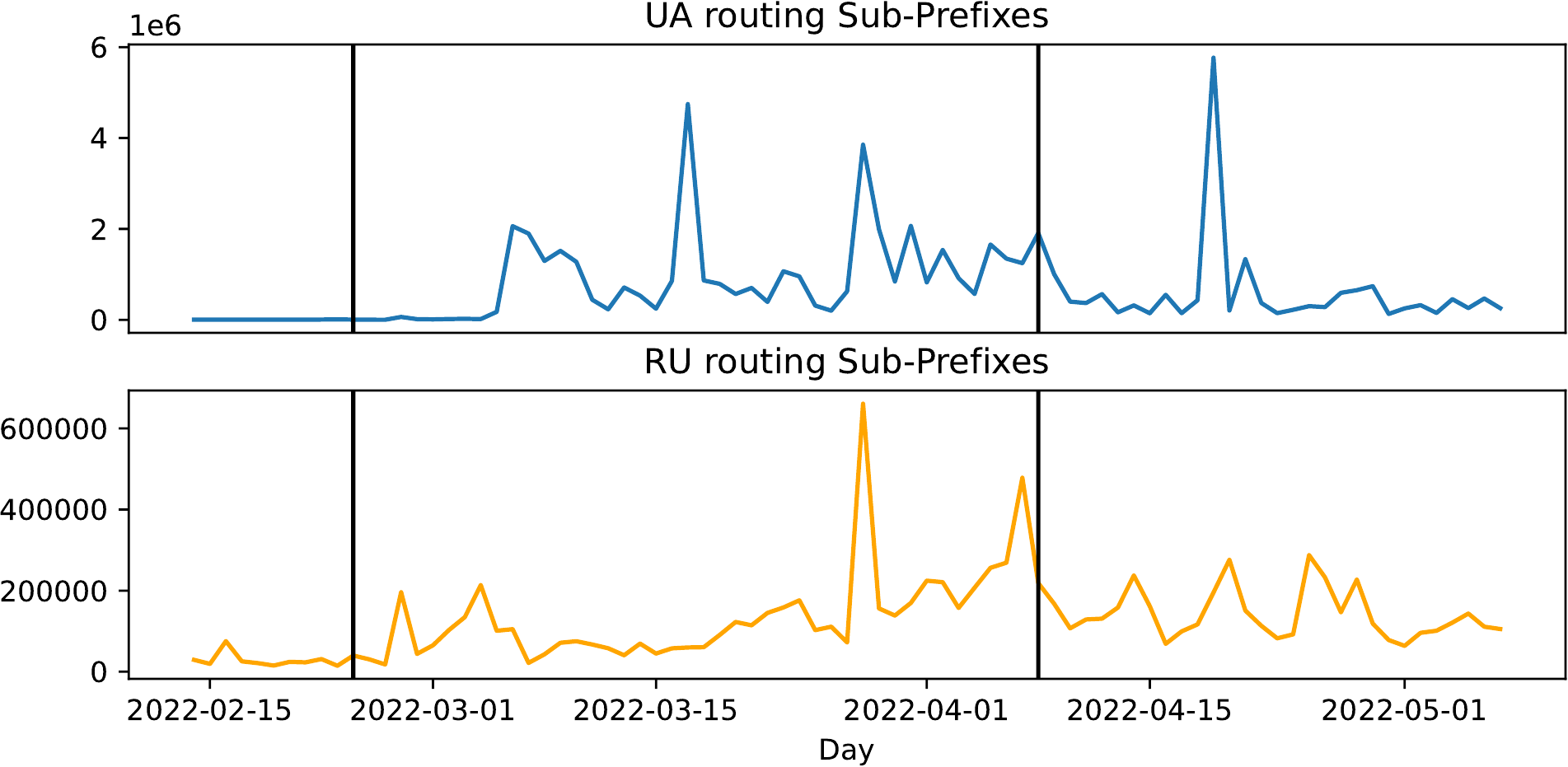}
    \caption {Number of BGP announcements showing ASes routing their sub-prefixes.}
    \label{fig:routing-subprefixes}
\end{figure}

Finally, we show an interesting phenomenon, which was already partially highlighted in the previous sections. In Figure~\ref{fig:routing-subprefixes}, we show the number of BGP announcements produced by Ukrainian and Russian ASes that started routing their own sub-prefixes. As can be observed, for both countries, the activity goes from almost zero to very high peaks, i.e., 6\,000\,000 announcements for Ukraine and 600\,000 announcements for Russia. This activity could be due to ASes that try to defend from hijack attacks, or Sub-Prefix and Its AS attacks carried out by other ASes that are neither Russian nor Ukrainian. However, as pointed out in~\cite{Almusawi2017:bgp}, BGP activity can be triggered also by malfunctions or attacks in other cyberdomains, such as worms or viruses spread.

\subsection{Discussion}
\label{subsec:discussion}
In this section, we provided a quantitative analysis of the phenomena that occurred in the Russian and Ukrainian Internet during the Russian invasion of Ukraine, from an inter-domain routing standpoint. We also provided some of the reasons that could explain these phenomena, however, explaining the true reasons behind them is almost impossible, as it would require being on site where and when the actions occur. This is obviously beyond the authors' possibilities and the scope of this paper. However, despite the hypothetical nature of a few considerations, we strongly believe that the value of the quantitative analysis underlying them provides an unprecedented measure of the conflict impact on the Internet.

\section{Impact on Latency}
\label{sec:latency}
We collected raw latency data about the Ukrainian Internet using RIPE Atlas \cite{staff2015ripe}.
Atlas is an Internet measurement platform managed by RIPE NCC, with an extensive presence in Europe and thus particularly suitable for observing the performance of the Internet in Ukraine. 

The results of Atlas measurements are collected and stored in the Atlas backend infrastructure. Such measurements have been frequently used in the past for monitoring the status of critical Internet infrastructure, such as DNSMON \cite{dnsmon}, or for research purposes, such as evaluating the impact of the mutated lifestyle due to the COVID pandemic on network performance \cite{CANDELA2020107495}. 
Internet performance is monitored using classical network tools. In particular, latency is observed using the ICMP-based version of ping. Each time a latency measure is triggered, a probe collects three RTT samples towards the considered target. 

\begin{figure*}[!t]
\centering
\subfloat[UA]{\includegraphics[width=0.3\textwidth]{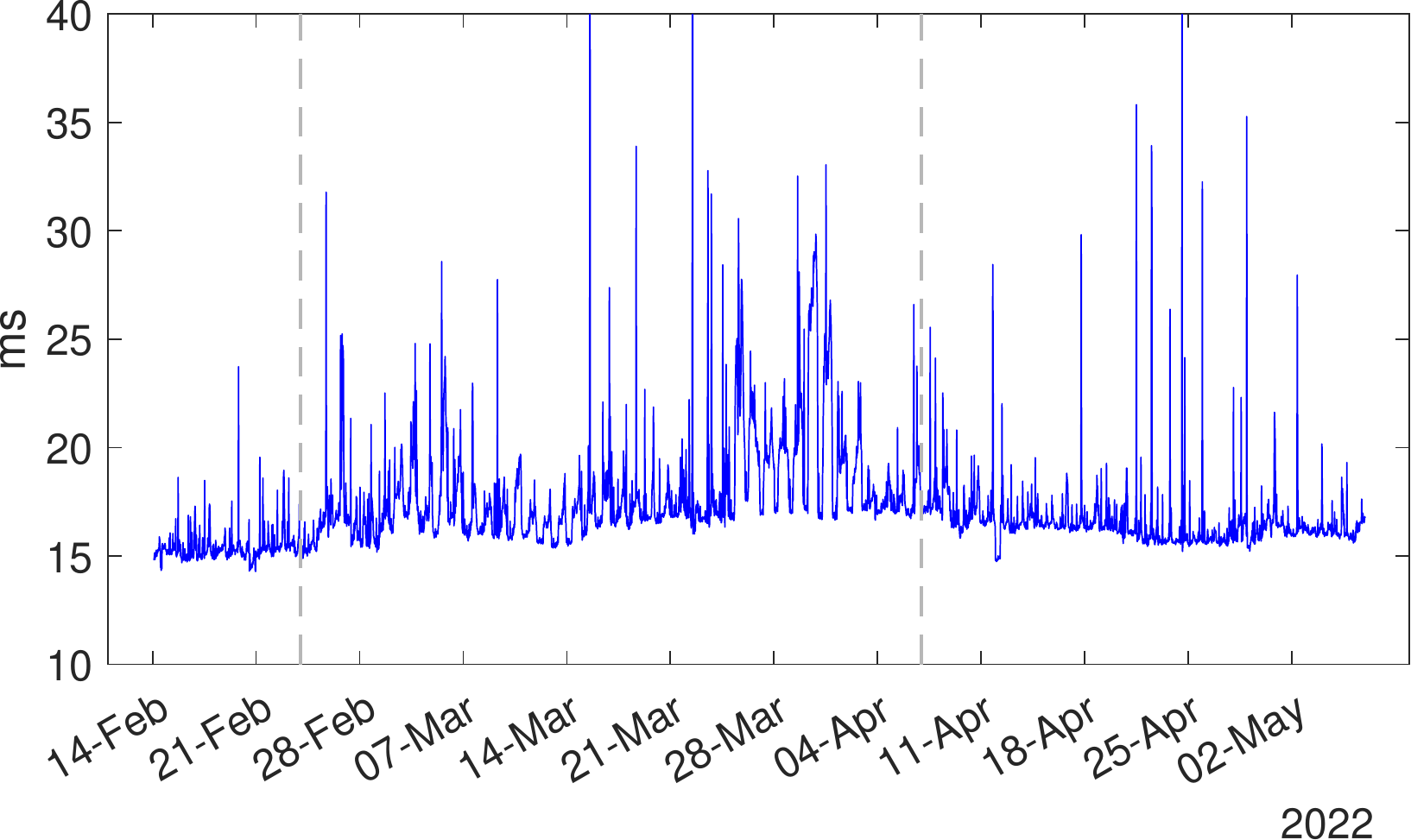}%
\label{fig:uaua}}
\hfil
\subfloat[RU]{\includegraphics[width=0.3\textwidth]{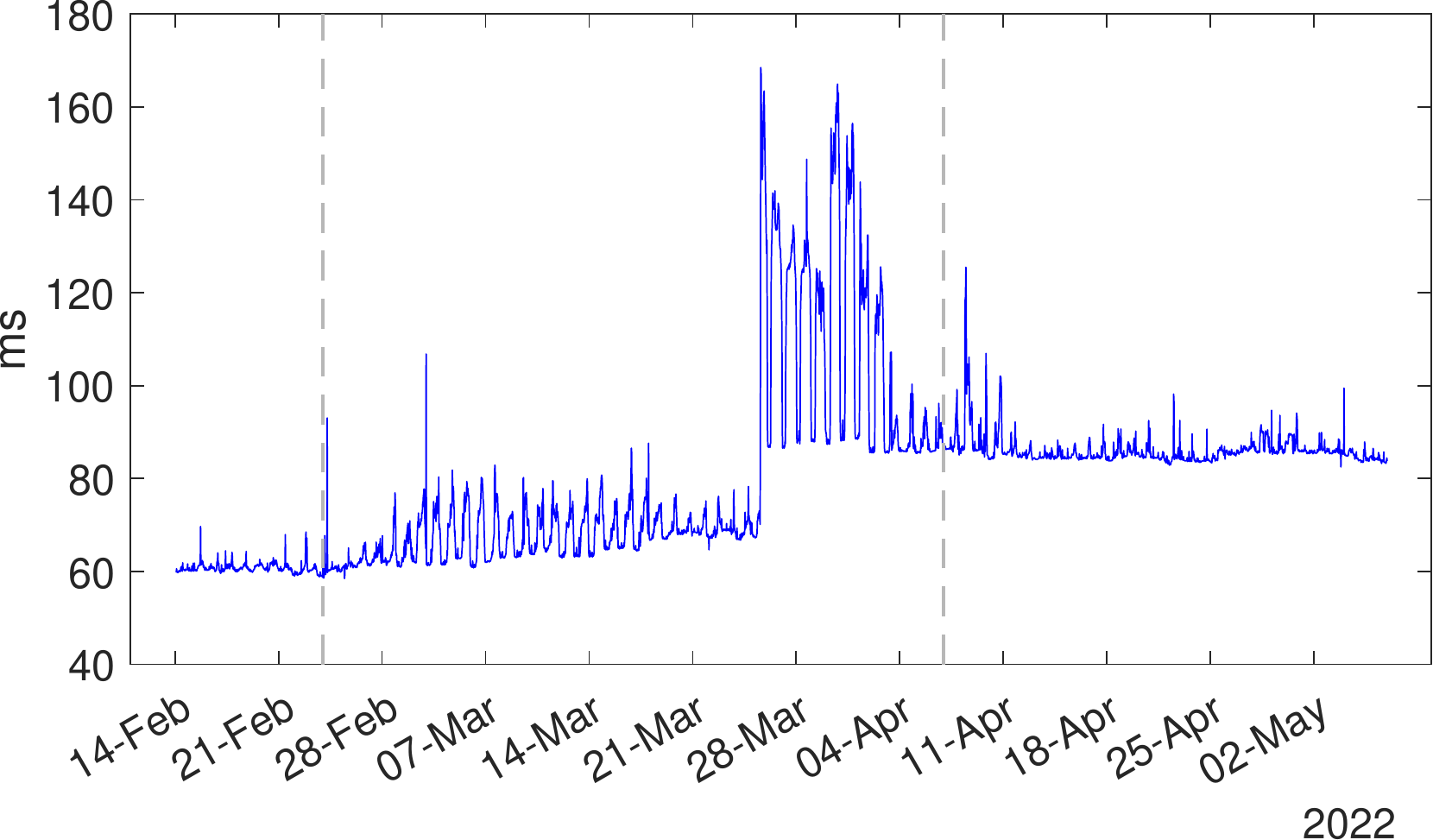}%
\label{fig:uaru}}
\hfil
\subfloat[Europe]{\includegraphics[width=0.3\textwidth]{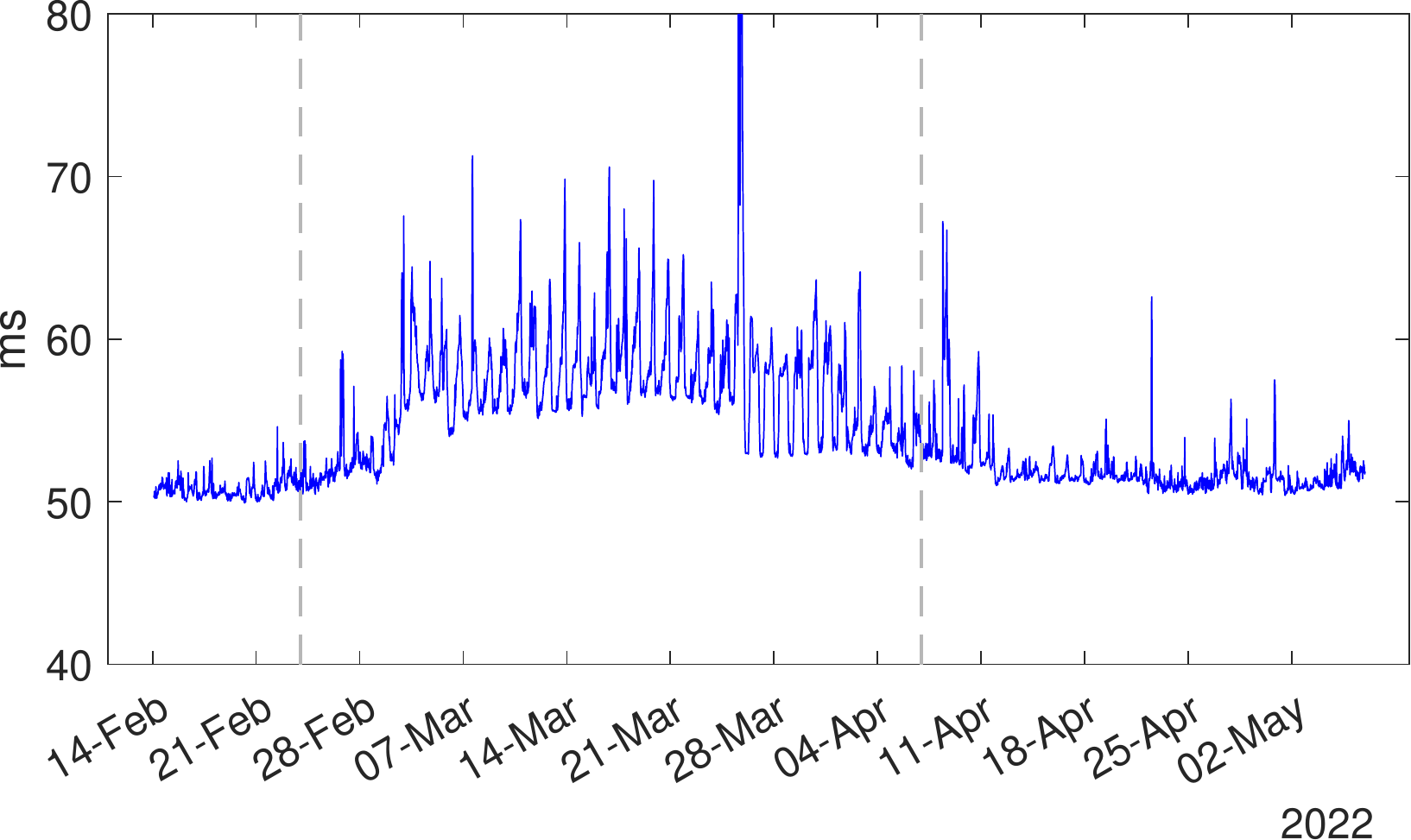}%
\label{fig:uaeu}}
\caption{Average latency, 30 min buckets, from UA-located probes to targets located in UA, RU, and Europe.}
\label{fig:latency}
\end{figure*}

AMs are particularly useful to observe the presence of possible variations in the performance of the Internet in a region because they are generally periodic. This allows comparing the performance of some networks at a given time to a stable baseline. Similarly to the routing analysis, we used the AMs falling in the period from 14 Feb 2022 to 7 May 2022, and the initial ten days are the baseline against which the performance during the war is compared. To avoid that some pairs of nodes participating in the measurements were available only during a small fraction of the considered period, we removed all the measurements originated by source-target pairs that were unavailable for more than 50\% of the period. The position of Atlas nodes is known, as it has to be provided by the hosting individual or organization. Latency values have been aggregated using bins with a duration of 30 minutes. In particular, we computed the average value of the three samples collected at each measurement attempt, and then computed the average for all the values falling in the same bin. Each different source-target pair has been included only once per bin, so that the result is not biased by pairs that produce more results than other pairs. 

We studied the impact of the war on the latency of the Internet in Ukraine when both source and destination are in Ukraine (UA $\rightarrow$ UA), and when the source is in Ukraine and the target is outside the country and located in Russia or Europe (UA $\rightarrow$ RU, UA $\rightarrow$ Europe). Figure~\ref{fig:latency} shows the ICMP-based latency for the three considered possible positions of the targets. In all cases, the latency increases significantly after the start of the war, identified by the first dashed line on the plots. The average latency after 24 Feb, is higher than the ones observed in the 10 days before the start of the conflict. The increase is +13\%, +35.5\%, and +7.8\% for the UA $\rightarrow$ UA, UA $\rightarrow$ RU, UA $\rightarrow$ Europe scenarios, respectively. The periodic peaks that are visible throughout the monitored period are related to the different loads imposed on the network by human-driven activities, higher during the day and lower during night-time. Peaks become more evident during the conflict, compared to the baseline period. The standard deviation of latency during the war is approximately 3.4 times the one observed before the conflict for the UA $\rightarrow$ UA scenario. The increase is even larger for the other two scenarios: 15.4 for UA $\rightarrow$ RU and 6.7 times for UA $\rightarrow$ Europe. It is interesting to notice that the degradation of performance for the UA $\rightarrow$ RU scenario is worse than the UA $\rightarrow$ UA one. The plots also report a second dashed line, corresponding to a second phase of the conflict (7 Apr) when battles started concentrating in the southeast part of Ukraine. 

We also estimated the packet loss by counting the number of missing echo replies, again using 30 min bins. The packet loss increases significantly after the start of the war: +249\% for the UA $\rightarrow$ UA scenario, compared to the baseline period. 

\begin{figure}[!t]
    \centering
    \includegraphics[width=0.9\columnwidth]{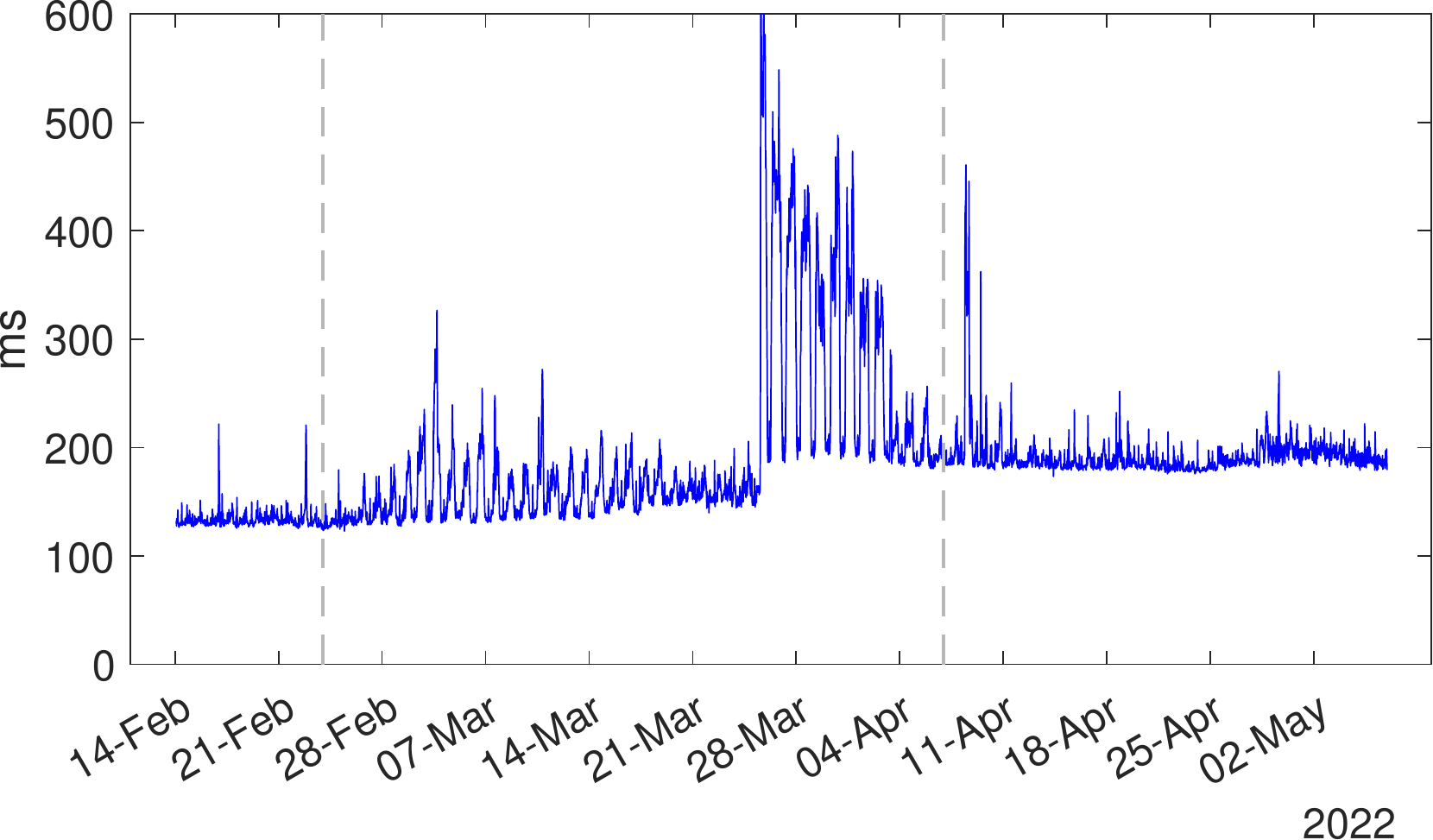}
    \caption {$UA \rightarrow RU$ average latency, HTTP, 30 min bins}.
    \label{fig:http}
\end{figure}
 
Finally, we studied the latency at the HTTP level, by relying on part of AMs carried out using such a protocol. Results are shown in Figure~\ref{fig:http} for the UA $\rightarrow$ RU scenario. Similarly to the other observed latency metrics, the HTTP-level latency becomes much higher and variable during the conflict (+44.2\% and +593\%, in terms of average and standard deviation). The bump in terms of BGP updates starting at the end of March and discussed in the previous sections is significantly overlapped with the period characterized by the largest fluctuations in HTTP latency. 
 
Besides increased variability, Figure~\ref{fig:uaru} shows a step-like trend when considering the minimum values of latency. To understand the causes of such an increase in the minimum latency, we analyzed the first 10 weeks of data as follows. First, we selected all the source-target pairs that include a step-like trend and then analyzed their paths at the AS level. In particular, we retrieved a set of traceroute measurements involving the selected source-target pairs and obtained the AS number for each IP address found in the path using RIPEstat. We also checked whether the IP addresses belonged to an IXP, using data extracted from PeeringDB~\cite{peeringdb}, or to a Tier 1 AS\footnote{A Tier 1 AS is an AS that contributes to international connectivity. The distinctive trait of Tier 1 ASes is that they are at the top of the hierarchy of the Internet AS-level graph. Tier 1 ASes, in other words, do not have any providers.}, using to this purpose the list provided in~\cite{abenrepo}. Then, we compared the first week of measurements with the last one, looking for ASes that disappeared from the paths or that started being involved. The AS that was abandoned more frequently was Megafon (RU, AS31133), followed by Dataline (UA, AS35297). We also observed that during the last week of measurements, some IXPs were not included anymore in the paths of the selected source-target pairs. The IXPs that were not present anymore are MSK-IX (RU), DE-CIX (DE), and DTEL-IX (UA), in order of decreasing frequency. At the same time, some IXPs started being used, in particular NL-IX (NL). The overall decreasing number of paths involving an IXP was accompanied by a larger adoption of Tier-1s, in particular Lumen (AS3356), Cogent (AS174), Arelion (AS1299). In many cases, the paths that showed the step-like trend in the end-to-end latency appear to have as a common factor a partial shift from peering to transit, thus involving companies located higher in the Internet hierarchy. To further confirm this analysis of the phenomenon, we computed the average rank and the average customer cone of all the ASes found in the paths using the data provided by ASRank~\cite{caida_asrank}. In ASRank, the AS with the largest customer cone is given rank 1. The customer cone of an AS is the set of ASes that can be reached from such AS when following only provider-to-customer links in the AS graph. The average rank during the first week of measurements (before the start of the conflict) was 400, whereas the average size of the customer cone was 1\,272. During the last week of measurements, the two values were 256 and 8\,461, respectively. 
 
 \begin{figure}[!t]
    \centering
    \includegraphics[width=\columnwidth]{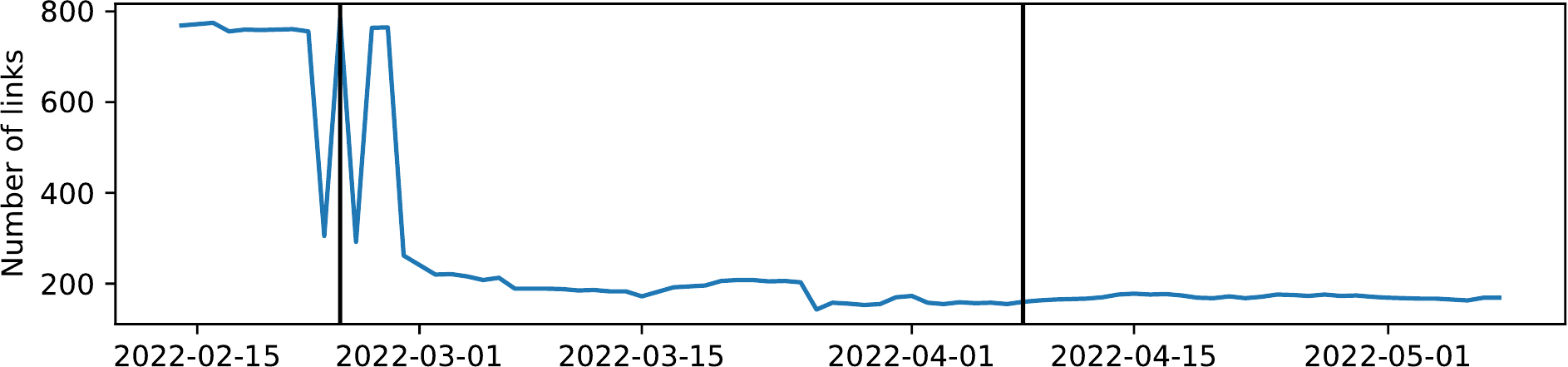}
    \caption {Number of links in the Ukrainian AS-level topology that directly connect Ukrainian ASes to Russian ASes.}.
    \label{fig:uarulinks}
\end{figure}
 
We extended this analysis by computing the Internet topology as seen from the Ukrainian perspective during the whole observation period ($\sim12$ weeks). To do so, we used the software in~\cite{abenrepo}. For each Ukrainian AS, the software collects all its neighbors from RIPEstat, which in turn extracts this information from BGP data collected by RIPE RIS. The Ukrainian topology was collected at daily snapshots, and, for each day, we computed the number of direct links between Ukrainian and Russian ASes. The results are depicted in Figure~\ref{fig:uarulinks}. As can be noticed, the number of links connecting Ukrainian and Russian ASes experiences a sudden drop, from almost 800 to approximately 200, after some days of instability, right after the start of the war. We then compared the first week of measurements with the last one. We collected all the direct links between Ukrainian and Russian ASes in the first week and in the last week. From these links we extracted the ASes involved, divided by country, Russia and Ukraine. For each AS, we then computed how many links it establishes. The number of Ukrainian ASes establishing links with Russian ASes is 30 in the first week and 19 in the last week, which means that approximately 30\% of Ukrainian ASes interrupted their direct connections with Russian ASes. The number of Russian ASes is instead 523 in the first week and 22 in the last week, thus just 4\% of ASes are still directly connecting with Ukrainian ASes. For these two sets of ASes, we then computed the average customer cone size, for the first and the last week of measurements. In the first week, the average customer cone size for Ukrainian ASes directly connected to Russian ASes is 105, while in the last week it is 160. For Russian ASes, the average customer cone size is 51 in the first week, and 912 in the last week. This means that just the biggest ASes, in terms of customer cone size, are still maintaining a direct connection with ASes of the other country. This is particularly evident for Russian ASes. It must be noticed that the Ukrainian AS that was establishing most of the links with Russian ASes in the first week is Uacity (AS48919), with 481 links. In the last week none of these links is still active.

\section{Conclusion}
The Ukraine-Russia conflict is one of the major catastrophic events occurring to a large country since when the Internet started playing a fundamental role in our society. The Internet is not only the cornerstone of the business and communications sectors, but also a key medium for individuals who need to access news in difficult times. According to ITU data, 75\% of the Ukraine population has access to the Internet, compared to 63\% of the world population~\cite{itu}. For Russia the same indicator has a value of 83\%. This shows how important is the Internet in such societies and economies. 

Our analysis provides a quantitative estimation of the impact of the conflict on the Ukrainian Internet. Data shows an intense rise of activity from the inter-domain routing standpoint, together with some suspicious activity that could be ascribed to cyberattacks. Some Ukrainian ASes got disconnected from the Internet as the Russian invasion proceeded, but in the last period of the observation they seem to start getting back online. The war activities impacted also other Internet-related aspects, with increased latency and an AS-level topology reconfiguration which saw Russian and Ukrainian ASes cease their direct connections. Despite the intense rise in BGP activity and the increased latency, the Ukrainian Internet proved to be quite resilient: considering the probes that were connected during the first week of the monitored period, approximately 83\% of them were connected also during the last week.

\section*{Acknowledgment}

This work is partially funded by the Italian Ministry of University and Research (MUR) in the framework of the CrossLab project (Departments of Excellence). We thank Massimo Candela, Ph.D. for the precious comments and suggestions. The
views expressed are solely those of the authors.

\bibliographystyle{elsarticle-num}

\begin{thebibliography}{10}
\expandafter\ifx\csname url\endcsname\relax
  \def\url#1{\texttt{#1}}\fi
\expandafter\ifx\csname urlprefix\endcsname\relax\def\urlprefix{URL }\fi
\expandafter\ifx\csname href\endcsname\relax
  \def\href#1#2{#2} \def\path#1{#1}\fi

\bibitem{9142776}
R.~Fontugne, K.~Ermoshina, E.~Aben, The internet in crimea: a case study on
  routing interregnum, in: 2020 IFIP Networking Conference (Networking), 2020,
  pp. 809--814.

\bibitem{donbass21}
K.~Limonier, F.~Douzet, L.~Pétiniaud, L.~Salamatian, K.~Salamatian, {Mapping
  the routes of the Internet for geopolitics: The case of Eastern Ukraine},
  First Monday 26~(5) (2021).

\bibitem{graham-cumming_2022}
J.~Graham-Cumming, {Internet traffic patterns in Ukraine since February 21,
  2022},
  \url{https://blog.cloudflare.com/internet-traffic-patterns-in-ukraine-since-february-21-2022/},
  accessed: 2022-08-11 (Mar 2022).

\bibitem{prince_2022}
M.~Prince, {Steps we've taken around Cloudflare's services in Ukraine, Belarus,
  and Russia},
  \url{https://blog.cloudflare.com/steps-taken-around-cloudflares-services-in-ukraine-belarus-and-russia/},
  accessed: 2022-08-11 (Mar 2022).

\bibitem{siddiqui_2022}
A.~Siddiqui, {Did Ukraine suffer a BGP hijack and how can networks protect
  themselves?},
  \url{https://www.manrs.org/2022/03/did-ukraine-suffer-a-bgp-hijack-and-how-can-networks-protect-themselves},
  accessed: 2022-08-11 (Mar 2022).

\bibitem{aben}
E.~Aben, The resilience of the internet in ukraine,
  \url{https://labs.ripe.net/author/emileaben/the-resilience-of-the-internet-in-ukraine/}.

\bibitem{10.1145/2079360.2079362}
K.~Cho, C.~Pelsser, R.~Bush, Y.~Won, {The Japan Earthquake: The Impact on
  Traffic and Routing Observed by a Local ISP}, in: Proceedings of the Special
  Workshop on Internet and Disasters, SWID '11, Association for Computing
  Machinery, 2011.

\bibitem{Heidemann12d}
J.~Heidemann, L.~Quan, Y.~Pradkin, {A Preliminary Analysis of Network Outages
  During Hurricane Sandy}, Tech. Rep. ISI-TR-2008-685b, USC/Information
  Sciences Institute (2012).

\bibitem{FAVALE2020107290}
T.~Favale, F.~Soro, M.~Trevisan, I.~Drago, M.~Mellia, {Campus traffic and
  e-Learning during COVID-19 pandemic}, Computer Networks 176 (2020) 107290.

\bibitem{CANDELA2020107495}
M.~Candela, V.~Luconi, A.~Vecchio, {Impact of the COVID-19 pandemic on the
  Internet latency: A large-scale study}, Computer Networks 182 (2020) 107495.

\bibitem{zhuo2011egypt}
X.~Zhuo, B.~Wellman, J.~Yu, {Egypt: The first internet revolt?}, Peace magazine
  27~(3) (2011) 6--10.

\bibitem{doi:10.1177/1748048512459147}
J.~Groshek, {Forecasting and observing: A cross-methodological consideration of
  Internet and mobile phone diffusion in the Egyptian revolt}, International
  Communication Gazette 74~(8) (2012) 750--768.

\bibitem{10.1145/3452296.3472916}
S.~A. Jyothi, {Solar Superstorms: Planning for an Internet Apocalypse}, in:
  Proceedings of the 2021 ACM SIGCOMM 2021 Conference, Association for
  Computing Machinery, 2021, p. 692–704.

\bibitem{modelling}
E.~K. {\c C}etinkaya, D.~Broyles, A.~Dandekar, S.~Srinivasan, J.~P.~G.
  Sterbenz, {Modelling communication network challenges for Future Internet
  resilience, survivability, and disruption tolerance: a simulation-based
  approach}, Telecommunication Systems 52~(2) (2013) 751--766.

\bibitem{ripencc}
{RIPE Network Coordination Center}, \url{https://www.ripe.net/}, accessed:
  2022-08-11.

\bibitem{riperis}
{RIPE NCC Routing Information Base},
  \url{https://www.ripe.net/analyse/internet-measurements/routing-information-service-ris},
  accessed: 2022-07-20.

\bibitem{bgprfc}
Y.~Rekhter, S.~Hares, T.~Li, {A Border Gateway Protocol 4 (BGP-4)}, RFC 4271
  (2006).

\bibitem{bgpstream}
{BGPStream}, \url{https://bgpstream.caida.org/}, accessed: 2022-07-20.

\bibitem{ripestat}
{RIPEstat}, \url{https://stat.ripe.net/}, accessed: 2022-07-20.

\bibitem{staff2015ripe}
{Staff, RIPE NCC}, {Ripe atlas: A global internet measurement network},
  Internet Protocol Journal 18~(3) (2015).

\bibitem{wikibattles}
{Timeline of the 2022 Russian invasion of Ukraine},
  \url{https://en.wikipedia.org/wiki/Timeline_of_the_2022_Russian_invasion_of_Ukraine},
  accessed: 2022-07-20.

\bibitem{Almusawi2017:bgp}
B.~Al-Musawi, P.~Branch, G.~Armitage, {BGP Anomaly Detection Techniques: A
  Survey}, IEEE Communications Surveys \& Tutorials 19~(1) (2017) 377--396.

\bibitem{dnsmon}
C.~Amin, M.~Candela, D.~Karrenberg, R.~Kisteleki, A.~Strikos, {Visualization
  and Monitoring for the Identification and Analysis of {DNS} Issues}, in:
  Proceedings of the Tenth International Conference on Internet Monitoring and
  Protection, 2015.

\bibitem{peeringdb}
{PeeringDB}, \url{https://www.peeringdb.com/}, accessed: 2022-08-10.

\bibitem{abenrepo}
{Visualising AS Hegemony per country},
  \url{https://github.com/InternetHealthReport/country-as-hegemony-viz},
  accessed: 2022-08-11.

\bibitem{caida_asrank}
{CAIDA AS Rank}, \url{http://as-rank.caida.org}, accessed: 2022-07-20.

\bibitem{itu}
{ITU DataHub}, \url{https://datahub.itu.int/}, accessed: 2022-08-10.

\end{thebibliography}

\end{document}